\documentclass[a4paper]{spie}  

\usepackage[]{graphicx}
\usepackage{amssymb,amsmath,amsfonts}

\title{Photonics-based mid-infrared interferometry: 4-year results of the ALSI project and future prospects }

\author{Lucas~Labadie\supit{a}, 
Stefano~Minardi\supit{b},
Jan~Tepper\supit{a},
Romina~Diener\supit{c},
Balaji Muthusubramanian\supit{a},
Jorg-\"Uwe~Pott\supit{d},
Stefan Nolte\supit{c},
Simon~Gross\supit{e},
Alexander~Arriola\supit{e},
Michael~J.~Withford\supit{e}
\skiplinehalf
\supit{a} I.\,Physikalisches Institut, Universit\"at zu K\"oln, Z\"ulpicher Str. 77, 50937 K\"oln, Germany \\
\supit{b} Leibniz-Institut f\"ur Astrophysik Potsdam, An der Sternwarte 16, D-14482 Potsdam, Germany\\
\supit{c} Institute Appl. Phys., Friedrich-Schiller-Universit\"at, Max-Wien-Pl. 1, 07743 Jena, Germany\\
\supit{d} Max Planck Institut f\"ur Astronomie, K\"onigstuhl 17, 69117 Heidelberg, Germany\\
\supit{e} MQ Photonics Research Centre, Department of Physics and Astronomy, Macquarie University, NSW 2109, Australia
}

\authorinfo{Further author information: (Send correspondence to L.Labadie)\\L.Labadie: E-mail: labadie@ph1.uni-koeln.de, Telephone:\,+49 221 470 3493
}

\begin{document} 
  \maketitle 

\begin{abstract}
In this contribution, we review the results of the ALSI project (Advanced Laser-writing for Stellar Interferometry), aimed at assessing the potential of ultrafast laser writing to fabricate mid-infared integrated optics (IO) devices with performance compatible with an implementation in real interferometric instruments like Hi5 or PFI. Waveguides for the L, L$^{\prime}$ and M bands with moderate propagation losses were manufactured in Gallium Lanthanum Sulfide and ZBLAN glasses and used to develop photonic building blocks as well as a full mid-IR 4-telescope beam combiner. We discuss the advantages and disadvantages of the tested combiners and discuss a possible roadmap for the continuation of this work.\end{abstract}



\section{Introduction and goals of the ALSI project}

Observations at high-angular resolutions in the mid-infrared are at high demand in the community for the study of distant AGNs, protoplanetary disks and young planets or protoplanets. From the ground, the 3--5\,$\mu$m spectral range is a sweet spot for this type of interferometrc observations, and it will be in part covered with the MATISSE instrument at the VLTI in the near future. 
Building on the excellent heritage of integrated optics beam combiners in Pionier and Gravity, the ALSI -- Advanced Laser-writing for Stellar Interferometry -- project was started mid-2014 to strengthen the route of mid-infrared integrated optics for interferometry developed with the Ultrafast Laser Writing (ULI) platform. The ULI approach had demonstrated by that time promising results, both for proof-of-concept devices\cite{Rodenas2012} and on-sky demonstrators\cite{Jovanovic2012}, and appeared as a valid alternative to lithographic techniques classically used for devices operating around 1.55\,$\mu$m\cite{Lebouquin2011}\,. In particular, one identified goal was to advance towards a deeper understanding of the achievable performances in relation to the stringent requirements for long-baseline interferometry (stable transfer function, sufficient throughput, minimum spectral bandwidth etc...). Finally, bridging the gap with near-IR integrated optics where 4-telescope beam combiners are well established was the ultimate goal of ALSI. The manufacturing of IO-devices is tested in infrared glasses such as Gallium Lanthanum Sulphide (GLS) and ZBLAN, which transparency range encompasses the K--M bands. This property can also be exploited for dual-band groupe delay fringe-tracking\cite{Muthusu2016} in high extinction star forming regions.\\
We believe that the benefit for long-baseline interferometry could become very important in the context of interferometric instrumentation focusing on the characterization of giant planets and circumstellar environments around pre-main sequence and young main sequence stars: current projects such as PFI and Hi-5 are directly concerned. The future foresees an increasing number of telescopes for an "ALMA-like" optical interferometer, as well as a regain of interest for nulling interferometry, which remains the only technique capable of characterising the atmosphere of terrestrial planets around Sun-like stars\cite{Leger2015,Kammerer2018}\,.

\section{Instrumental requirements}

\subsection{Interferometric contrast}

The fidelity of the interferometric image reconstruction depends on the extent of the UV coverage and on our ability to probe a large range of calibrate visibility amplitudes corresponding to the low- and high-spatial frequencies. This means in return that the raw instrumental contrasts should be high ($\sim$0.8--0.9) and only mildly affected by photometric imbalance. When operating in broadband conditions a high visibility white-light interferogram must be measured as well.

\subsection{Propagation losses and total throughput}

Ideally, the highest possible throughput of the integrated optics beam combiner is sought to reach high sensitivity. The three main contributors to a decrease in throughput are: 1) material transparency and reflection losses; 2) intrinsic and extrinsic propagation losses;  3) bending losses. The first point requires to use infrared material bulks transparent in the 2--5\,$\mu$m range with attenuation coefficient as low as few 10$^{-3}$\,cm$^{-1}$\footnote{Infrared glasses/crystals like ZnSe, GLS or Lithium Fluoride have absorption coefficients of $\sim$5$\times$10$^{-4}$ to 5$\times$10$^{-3}$\,cm$^{-1}$ around 4\,$\mu$m, whereas Silica (SiO$_2$ goes down to $\sim$10$^{-5}$\,cm$^{-1}$ at 1\,$\mu$m.}. Infrared glasses tend to suffer from high Fresnel losses because of a high refractive index which is around $n\sim$2.4 at 3\,$\mu$m for the GLS. Anti-reflection coatings may in principle mitigate this effect. Points 2) and 3) depends on the level of field confinement on one hand and on the manufacturing process on the other hand. For  an ideal waveguide, a low field confinement results in larger propagation and bending losses than for high confinement assuming the same waveguide geometry. The field confinement is controlled by the core/cladding index contrast, with values of few 10$^{-3}$ to 10$^{-2}$ that appears feasible with the ULI platform in chalcogenide glasses\cite{Rodenas2012}\footnote{The Silica-based Gravity beam combiner has $\Delta$n=0.01}. Additional -- and potentially significant -- extrinsic losses may arise from the manufacturing process that creates impurities and scattering centers in the waveguide, depending on the laser fluences. Last but not least, injection losses from free space optics or fibers feeding the chip further degrade the transmission. 
The challenge resides in mitigating these effects to an "acceptable" level for on-sky usage: as a matter of comparison, the global throughput of the fibered Gravity beam combiner over the K band is about 55\% (Perraut et al. 2018, in press).

\subsection{Single-mode behavior}

The single-mode property of integrated optics for interferometry has been a central feature of this technology since the early days of its implementation\cite{Kern1997}\,.  The resulting modal filtering of the wavefronts guarantees a higher stability of the instrument transfer function and a better calibration of the interferometric visibilities, with a consequent reduction of the error bars. The implication of this requirement is the manufacturing of waveguides with relatively small cores -- typically on the order of $\sim$10\,$\mu$m at $\lambda$=3\,$\mu$m -- which in return implies more stringent conditions for coupling the telescope beam into the single-mode waveguide in comparison to the multimode one\cite{Ruilier2001}\,. Interestingly, studies on the trade-off between single-mode and multimode waveguides in the low flux conditions frequently encountered in astronomy have been conducted\cite{Tatulli2004}\,.


\subsection{Birefringence and polarisation mismatch}

Birefringence in the context of stellar interferometry is typically related to an alteration of the polarisation state of the propagating wave due to the waveguide properties. A description of the birefringence type and impact in this context is available elsewhere\cite{Labadie2016}\,. For the different arms of an interferometer, differential birefringence may result in polarisation mismatch that decreases the instrumental contrast following $V_{pol}$=2$\cos(\alpha)$/(1+$\cos(\alpha)$), where $\alpha$ is the angular mismatch. Polarisation issues are usually treated in interferometry by splitting the polarisation by mean of a Wollaston prism, though this approach might be disadvantageous in terms of sensitivity (50\% of the flux is lost or redirected) and of increased complexity of the optical setup in case polarimetry is not required scientifically.

\subsection{Chromatic dispersion}

Differential chromatic dispersion between different arms of IO beam combiner is seen in broadband operation in the form of a spreading of the white-light interferogram. This is interpreted as each constituting wavelength experiences a different location of its zero-OPD position, with the consequence of a decrease of the instrumental contrast and a strong chromaticity of the instrumental phase. Though this can be in principle calibrated, it needs to be mitigated to some level by using a manufacturing platform as stable as possible. In operation, the impact of this effect is reduced by spectrally dispersing the fringes, which is anyway a very important functionality from the astrophysical point of view. Chromatic dispersion has been studied in other works\cite{Foresto1995,Tepper2017a}\,.

\subsection{Compatibility with cryogenic operation}

At wavelengths $\lambda$$>$2\,$\mu$m the significant background due to the sky and instrument thermal emission may hamper any high-sensitivity measurement. The classical solution in the mid-infrared consists in baffling and cooling down as much as possible the environment seen by the detector. As the telescope itself cannot be cooled down, the instrument is usually placed in a cryogenic enclosure. The small footprint of integrated optics instrument relaxes in principle the constraints for such operation, but one has to assess first that the integrated optics component does not see its performance degraded by the low temperatures. This is even more a concern when fibers are connected to the component, as the contact point also suffers from aging at low temperatures. In Gravity, the fibered IO component is enclosed in the cryostat and does not appear to underperform.

\subsection{Importance of static measurement}

In modern interferometric instruments it is advantageous to obtain a static measurement of the fringe modulation, where no moving part is involved, and this for all the available baselines simultaneously. This is the principle behind the integrated optics ABCD beam combiner of Gravity, as well as of the DBC or multiaxial combination schemes capable of delivering instantaneous measurement of the visibilities, as opposed to the temporal encoding of the fringes used in MIDI/VLTI\cite{Leinert2003}\,. A significant advantage is that temporal biases in the signal acquisition are avoided. Hence, we focused in the ALSI project on the development of mid-infrared static beam combiners.

\section{Visibility-to-Pixel Matrix}\label{v2pm}

\begin{figure}[b]
\centering
\includegraphics[width=0.75\textwidth]{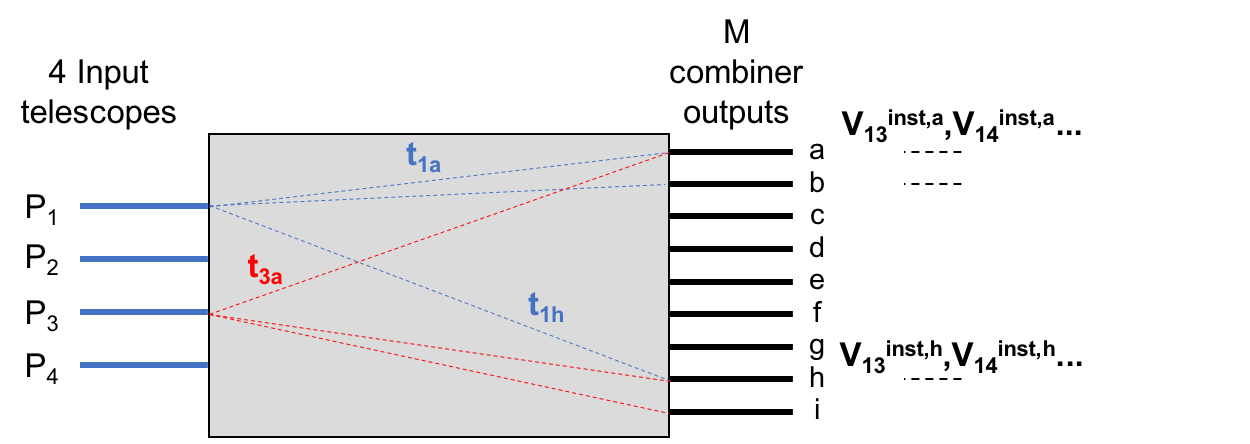}
\caption{Schematic illustration of the quantities involved to establish the V2PM matrix. $P_i$ are the input photometries, $t_{kj}$ are the terms of the $\kappa$-matrix, $V_{lm}$ and $\phi_{lm}$ are the visibility and phase terms.}\label{v2pm}
\vspace{0.0cm}
\end{figure}

Working with static beam combiner allows us to use the formalism of the V2PM\footnote{Visibility-to-Pixel-Matrix} matrix and its (pseudo-)inverse P2VM matrix. 
The instrumental behaviour of a N-telescope beam combiner can be completely characterised by the so-called V2PM matrix, which gives an algebraic relationship between the fluxes measured at the $M$ available outputs on one side, and the $N(N-1)/2$ complex visibilities and $N$ photometric measurements of an astrophysical source on the other side. 
The formalism has been addressed in several papers\cite{Tatulli2006,Lacour2008,Benisty2009} to which the reader can refer, hence we only briefly remind here some aspects relevant to this paper.\\
For a 4-telescope static beam combiner ($N$=4), the monochromatic flux measured at a given output $j$ can be related to the state of interference between the four input beams (1, 2, 3, 4) following (see Fig.~\ref{v2pmeq})
\begin{eqnarray}
I_j=\sum_{k=1}^N P_k.t_{kj} + 2\sum_{l=1}^{N-1}\mathop{\sum_{m>l}^N}  V_{lm}^{obj}V_{lm}^{inst,j} \sqrt{P_lP_m t_{lj}t_{mj}}\cos(\phi^{obj}_{lm})\cos(\phi^{inst,j}_{lm}) \nonumber \\
- 2\sum_{l=1}^{N-1}\mathop{\sum_{m>l}^N}  V_{lm}^{obj}V_{lm}^{inst,j} \sqrt{P_lP_m t_{lj}t_{mj}}\sin(\phi^{obj}_{lm})\sin(\phi^{inst,j}_{lm})\label{v2pmeq}
\end{eqnarray}

\noindent $P_k$ is the flux arising from telescope $k$ on the input $k$ of the beam combiner, $t_{kj}$ is the transmission term -- including Fresnel and propagation losses -- from telescope $k$ to output $j$ and form the terms of the so-called $\kappa$-matrix (see Fig.~\ref{v2pm}). $V^{obj}_{lm}$ and $\phi^{obj}_{lm}$ are, respectively, the intrinsic amplitude and phase of the complex visibility of the astrophysical object as measured with the baseline ($l$,$m$). $V^{inst,j}_{lm}$ and $\phi^{inst,j}_{lm}$ are, respectively, the instrumental visibility amplitude and phase at output $j$ for the baseline ($l$,$m$) measured for a point-like source ($V$=1, $\phi$=0\,rad). These terms fully characterise the instrumental response. Eq.~\ref{v2pmeq} can easily be expressed in the form of a matrix multiplication {\bf I=M$\times$V} with {\bf I} and {\bf V} the intensity vector {\bf I}=[$I_1$,$I_2$...$I_M$] and the visibility vector {\bf V}=[$P_1$,$P_2$, $P_3$, $P_4$,  2$V_{12}^{obj}\sqrt{P_1P_2}\cos(\phi^{obj}_{12})$, 2$V_{12}^{obj}\sqrt{P_1P_2}\sin(\phi^{obj}_{12})$ ... 2$V_{34}^{obj}\sqrt{P_3P_4}\cos(\phi^{obj}_{34})$,2$V_{34}^{obj}\sqrt{P_3P_4}\sin(\phi^{obj}_{34})$], respectively.\\
\\
For a 4-telescope interferometer with 6 baselines, the vector {\bf V} has 16 coordinates, i.e. 1 photometry term/telescope and 2 coherence terms/baseline. The vector {\bf I} has, respectively, 23 and 24 components for a 4-telescope DBC and a 4-telescope ABCD combiner. The matrix {\bf M} -- or V2PM -- is composed of the $\kappa$-matrix terms and the instrumental coherence terms. It is hence a 23$\times$16 and 24$\times$16 rectangular matrix for the two types of beam combiner, respectively. The 4-telescope pairwise static ABCD\cite{Benisty2009} is a special case of this formalism since only one baseline is encoded at one of the 24 available outputs, hence only two terms of the $\kappa$-matrix are present in the corresponding line of the V2PM, the others being set to zero in absence of cross-talks. Finally, in the case of the 2-telescope ABCD combiner studied in the ALSI project, the V2PM reduces to a 4$\times$4 matrix. Once the V2PM is established, it can be (pseudo-)inversed into the P2VM, with {\bf V} that can be retrieved through the {\bf V}= P2VM$\times${\bf I}. \\
\\
Both for the ABCD and DBC beam combiner, a regular calibration of the V2PM will be needed at the telescope, for instance prior to each observing night as this is done for the AMBER instrument. This calibration stage of the V2PM accounts for the transfer function for a given telescope configuration, as well as for the procedure of signal extraction and measurement, which has to be identical for the V2PM, the calibrator and the astrophysical sources.

\section{Device fabrication and key parameters}\label{DeviceFab}

\subsection{Principle of Ultrafast Laser Inscription}

The ALSI project has specifically focused on using the ULI technique as a manufacturing platform to develop mid-infrared IO devices in GLS substrates suitable for interferometric applications. Thanks to a collaborative work with the Macquarie University in Australia, we also explored the fabrication and performance of laser-written ZBLAN couplers. In summary, the ultrafast laser inscription technique allows localised and permanent change of the refractive index in a dielectric substrate by tightly focusing femtosecond pulses from a pulsed laser. Under high irradiance conditions the range of non-linearity of the refractive index has to be considered, inducing multi-photon absorption of the pulses. By translating the laser beam in three dimension in the substrate, 3D guiding structures can be obtained. The size and shape of the modified area depends on parameters such as pulse length, repetition rate, laser fluence... A detailed literature is available on the topic\cite{He2014,Gross2015}\,. Laser-writing has been demonstrated in planar substrates using CW laser\cite{Ho2006,Labadie2011}\,, but this approach lacks the 3D versatility of the ULI platform. Other groups develop promising approaches based on femtosecond-laser writing to develop sub-diffraction-limited structures in glass\cite{Lee2017}\,.

\subsection{Waveguide cross-section, achievable $\Delta$n, writing depth}\label{crosssection}

\begin{figure}[t]
\centering
\includegraphics[width=0.75\textwidth]{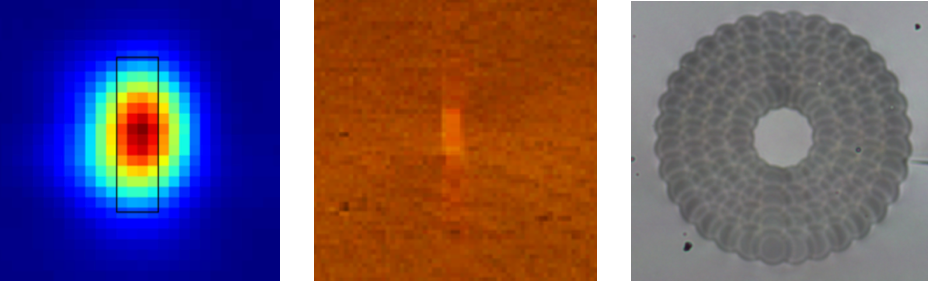}
\caption{Left: near-field imaging of the fundamental mode for a rectangular GLS waveguide with a 29$\times$22\,$\mu$m 1/e$^2$ MFD. The black contour is the footprint of the cross-section. Center: microscope image of the GLS rectangular waveguide (courtesy R. Diener). Right: circular core of a ZBLAN laser-written waveguide with a diameter of $\sim$40\,$\mu$m (bright central core).}\label{crosssec}
\vspace{0.0cm}
\end{figure}

The pair waveguide cross-section/index contrast is an important parameter to be considered in the design and fabrication process of a guiding structure. It determines, for instance, the single-mode/multimode behaviour, or the degree of field confinement in the waveguide core that constrains in return the amount of bending losses. 
Depending on the repetition rate of the femtosecond laser, guiding structures can be written in the athermal or thermal regime (cf. Gross \& Withford (2015)\cite{Gross2015} for details). This directly impacts the geometry of the irradiated region. \\
In the athermal regime, ultrafast lasers with few kHz to few hundred kHz repetition rate will generally imprint a localized index contrast matching the intensity distribution of the writing beam. Usually they are employed with low NA ($\sim$0.6) microscope objectives, which access a longer vertical writing depth in the substrate because of the larger free working distance. 
In this regime, 
an elongated higher index core -- or line -- in the longitudinal direction is imprinted, which reflects the increased Rayleigh range of the beam at the waist position compared to the case of a high NA ($\sim$1.25) objective. 
In the multipass technique, the laser spot is translated periodically in the transverse direction in order to form a square or rectangular waveguide cross section formed of several juxtaposed lines. 
This approach was chosen for laser writing in GLS glasses using a 500\,kHz repetition rate ultrafast laser delivering $\sim$400\,fs pulses. An example of a single-mode waveguide cross-section written in GLS substrate and the corresponding fundamental mode field distribution at 3.39\,$\mu$m are shown in Fig.~\ref{crosssec}. Assuming a step index profile for the core, we estimated an index contrast between 3$\times$10$^{-3}$ and 4$\times$10$^{-3}$. \\
The thermal regime is reached with femtosecond lasers delivering lower energy pulses but at higher repetition rates of few MHz. Increasing the power density at the laser focus is achieved with higher NA objectives, which however limits the writing depth because of the shorter free working distance. In this regime, an index modification with isotropic circular cross-section index and dimensions larger than the laser spot can be obtained. In collaboration with the Macquarie University, ZBLAN-based singlemode waveguides with a circular core diameter of $\sim$40\,$\mu$m have been manufactured and tested. Differently to the GLS substrate, a depressed-cladding is laser written to surround the unaffected core. The achieved index contrast in this case is about $\Delta$n=6$\times$10$^{-4}$.\\
Typically the achieved core size and index contrast by ULI significantly depends on the repeatability of the writing parameters (which depend on the laser stability etc...) and the batch of substrate being used (R.~R.~Thomson, private communication). Hence. these parameters need to be re-calibrated for any new run and batch. The long-term repeatability is otherwise poor. 

\subsection{Birefringence and tensile stress}

In Diener et al. (2018)\cite{Diener2018} it was shown that the output diffraction pattern of a symmetrically excited linear array of coupled waveguides depends on the input -- horizontal or vertical -- polarization of the excitation field, although the array of waveguides was written in GLS glass using a circular polarisation of the laser. Such a birefringence appears to result from long range tensile stress of the laser spot over the surrounding region where the next-to-come waveguide has to be written. This effect is empirically mitigated thanks to a regular increase of the laser writing speed from waveguide \#\,1 to waveguide \#\,N in the array, which allows a more symmetric distribution of the tensile stress and lower birefringence. This approach was applied for the fabrication of the DBC static combiners used in ALSI. %
 \\
The GLS directional couplers 
were not compensated for this effect. As a consequence, birefringence was observed at 3.39\,$\mu$m in the form of a polarisation-dependent splitting ratio of the flux and a change from linear to elliptical polarisation state when the input polarisation was intermediate between the vertical and horizontal direction\footnote{The direction of vertical polarisation coincides here with the "depth" degree of freedom of the ULI platform.}\,. We measured nonetheless that the polarisation behaviour of the coupler is similar for both the left and right inputs, leading to small {\it differential} birefringence. Hence, for such couplers the effect of the birefringence is estimated to result in less than 1.5--2\% loss in contrast. 
The effect of the shape birefringence resulting from a non circular cross-section of the waveguides in the GLS components has not been investigated here, but this is likely not to be a dominant effect in our work.

\subsection{Manufactured devices}

\subsubsection{Directional and asymmetric couplers}\label{dircouplers}

\begin{figure}[t]
\centering
\includegraphics[width=1\textwidth]{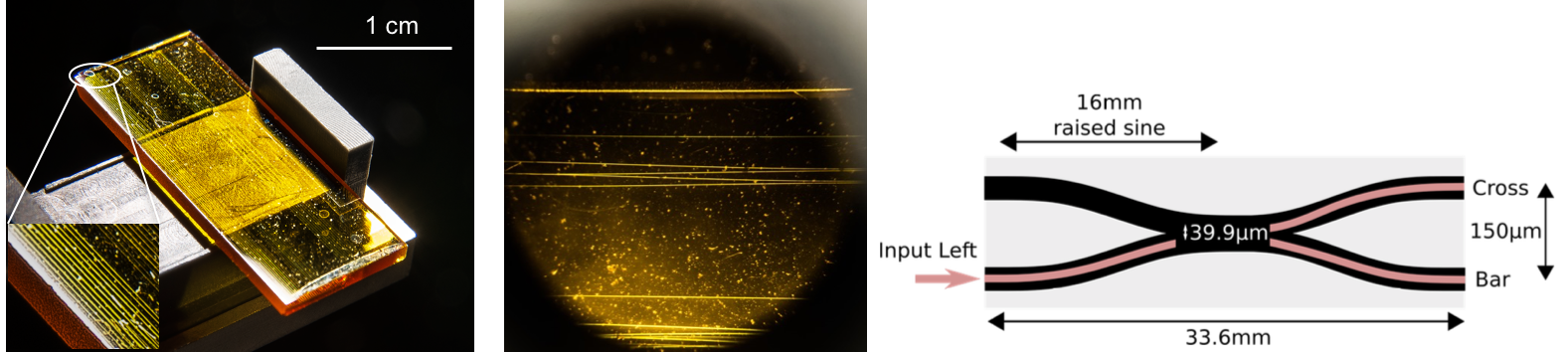}
\caption{Left and middle: Photo of a chip containing directional couplers fabricated in GLS glass and close view on the individual laser-written waveguides. Right: Design of the ZBLAN coupler characterised in the L$^{\prime}$ band\cite{Tepper2017b}.}\label{closeview}
\vspace{0.0cm}
\end{figure}

As a first brick for more elaborated devices, simple 2-telescope directional couplers were designed, manufactured and tested. Depending on the separation between the waveguides and on the interaction length, a 50/50 splitter can be designed at a given wavelength and be reversely used as a beam combiner. A important property of a directional coupler used as interferometric beam combiner is that the interferograms recorded at each output are in phase opposition as a result from energy conservation. 
We used the laser writing facility in Jena composed of a Yb:KGW laser at 1023\,nm delivering 400\,fs pulses at 500 kHz repetition rate to manufacture an integrated optics chip in GLS, a substrate with a refractive index n$\sim$2.3 at 3$\mu$m and a high transparency from 0.5 to 9 $\mu$m. 
Different designs were studied: the couplers are formed of S-bend waveguides with an amplitude of 50 and 75$\mu$m, respectively, to assess the impact of bending losses. This resulted into a separation of 120 and 170\,$\mu$m, respectively, between the inputs. In the interaction region where coupling between the two waveguides occurs, the tested coupling lengths were 0, 2, and 4\,mm. Finally, the separation between the two waveguides in the interaction region was varied from 20 to 22\,$\mu$m with 0.5\,$\mu$m spacing as this parameter greatly influences the coupling strength. In total 20 directional couplers sampling these different parameters were written and contained in a 25$\times$10$\times$1$\times$\,mm chip (see Fig.~\ref{closeview}). The waveguides were written at a 200\,$\mu$m depth below the surface of the substrate using a 0.6 NA objective and had a measured cross-section of $\sim$25$\times$10\,$\mu$m$^2$. The total length of a directional coupler was 25\,mm.\\
A similar coupler concept was fabricated in a ZBLAN substrate by writing a depressed cladding around the single-mode core of interacting waveguides. In this case the writing facility was composed of Ti:sapphire oscillator with a 5.1\,MHz repetition rate and $<$50\,fs pulses at 800\,nm. The lower index contrast implies writing smoother S-bend transitions to mitigate the bending losses, which consequently results in a longer coupler of 33.6\,mm for a separation of 150\,$\mu$m between the inputs/outputs (see Fig.~\ref{closeview}). 
In the ZBLAN directional coupler, the cores are brought to a center-to-center separation of 39.9\,$\mu$m, i.e. practically overlapping. The interaction distance is set to 0\,mm, meaning the waveguides are written to converge and directly depart from each others without sharing any straight interaction segment. \\
Both type of couplers are affected by intrinsic propagation losses depending on the degree of strong/weak confinement of the field, and by bending losses. Further extrinsic losses are typically present due to the presence of impurities and scattering centers.\\
A directional coupler in which the interaction length $d$ is an odd multiple of the beat length has an intrinsically chromatic splitting ratio.This is seen in the expression of the power coupled into one output $P_{\rm bar}$/$P_{\rm tot}$\,$\propto$\,$\sin^2$($Kd$/$\lambda$)\footnote{From Snyder \& Love (1983) "Optical Waveguide Theory", assuming a weak wavelength-dependence of the coupling.} where $\lambda$ is the wavelength and $K$ the coupling constant and $d$ the interaction length. Asymmetric couplers with a slight difference in the propagation constant $\Delta\beta$ between the two interacting waveguides can be implemented to flatten the splitting ratio over a limited bandwidth. This was tested on directional couplers for which the second segment in the interaction region was written at a slighter higher speed than for the first segment, with the goal of inducing the sought difference in the propagation constant. As for the directional couplers, several samples with an interaction length of $d$=0, 1, 2, 4\,mm where produced, and for each of them an increase of +0.5, +0.75, +1.5\,mm/s in the writing speed was set. Asymmetric couplers were only tested with the GLS platform and not with ZBLAN.

\subsubsection{ABCD units}

\begin{figure}[t]
\centering
\includegraphics[width=0.4\textwidth]{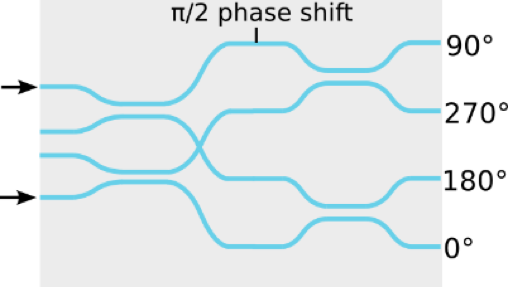}
\includegraphics[width=0.4\textwidth]{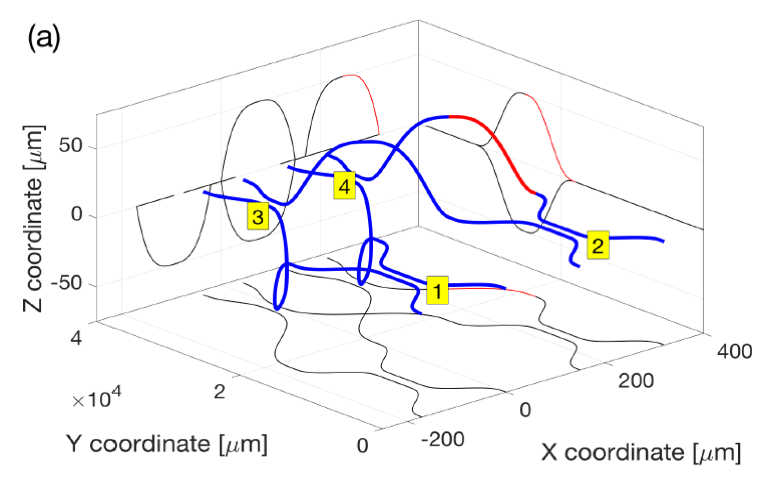}
\caption{Left: illustration of the 2-telescope ABCD device. The black arrows indicate the two inputs. A $\pi$/2 shift is introduced in the upper waveguide arm. Right: 3D ULI setup of the ALSI 2-telescope ABCD. The red portion indicates the region of altered propagation constant\cite{Diener2017}.}\label{abcd}
\vspace{0.0cm}
\end{figure}

In ALSI, we applied the know-how acquired with the directional coupler to develop more advanced functions like the 2-telescope ABCD unit. This unit block allows to sample a fringe sine wave in quadrature, i.e. in four different phase positions separated by $\pi$/2 with respect to each other. The ABCD values are recorded simultaneously, hence the visibility and phase estimators \cite{Colavita1999} can be derived in a static way. Such a functionality is implemented in the Gravity instrument\cite{Eisenhauer2017}\,. \\
A writing setup similar to the one used for the manufacturing of GLS coupler was used to manufacture several 2-telescope ABCD beam combiners. They are composed of cascaded couplers with a phase retardation of $\pi$/2 inserted in one of the arm by mean of a local variation of the writing speed along a small portion of the waveguide (see Fig.~\ref{abcd}). This induces a local variation of the propagation constant, which can be properly tuned to match the $\pi$/2 phase shift. The required index variation can be optimised with BeamProp. Technically, this was obtained by raising gradually the writing speed from 1\,mm/s to 1.4\,mm/s over a 2.5-mm-long section, kept constant for another 1.2\,mm and then decreased back to 1 mm/s within the final 2.5 mm. 
A circular polarization state was applied to the PHAROS femtosecond laser pulse by inserting a quarter-wave plate and a 0.35 NA objective was used to increase the vertical writing range. The benefit of the 3D ULI writing is ideally exploited here since any crossing between the waveguides is avoided, hence mitigating the risk of cross-talk typical of planar geometries. A sketch of the laser-written ABCD device is shown in Fig.~\ref{abcd} \cite{Diener2017}. Here, the challenge of the lab characterization resides in verifying if the device can operate in broadband -- or low spectral dispersion -- mode and still deliver high contrasts without significant chromatic dispersion (see Jocou (2014), Fig.12\cite{Jocou2014} for the Gravity beam combiner). This also implies that the injected $\pi$/2 shift must be relatively achromatic over the bandwidth of interest.

\subsubsection{4T-beam DBC combiners}

Directional couplers and ABCD units are the building blocks for 4-telescope Gravity-like beam combiners for the static derivation of the phase and visibility for the six baselines simultaneously. Since a large number of S-bend waveguides are involved in this type of architecture architecture, bending losses is a major issue in the overall perfromance of the final device in terms of throughput. We studied alternative solutions to ABCD-type beam combiners in the form of so-called Discrete Beam Combiners (DBC), which are solely composed of coupled channel waveguides. DBCs allow as well the static retrieval of the coherence function of the source for the N(N-1)/2 encoded baseline with the advantage that no bending losses are affecting the total throughput of the device. As we will see in Section~\ref{Labwork}, the current 2-telescope ABCD unit manufactured by ULI appears to exhibited a relatively low throughput due to poor field confinement and high bend losses. We investigated if the DBC approach based on channel waveguides could be more appropriate in our case as it was shown that the (theoretical) performances of DBCs in terms of sensitivity are comparable to ABCD-type beam combiners\cite{Minardi2016}\,. DBCs are normally composed of $(N+1)^2$ channel waveguides set in a two-dimensional array with respect to the propagation axis\cite{Minardi2012}, where $N$ is the number of telescope that are coherently combined. For ALSI, it was shown that a 23-waveguide lattice instead of 25 is still suitable for the interferometric combination of four telescopes. Among the $(N+1)^2$ available inputs, only a well-defined subset of $N$ waveguides to be fed with the beams of the $N$telescopes results into a V2PM with the lowest condition number. \\
A GLS-based Discrete Beam Combiner was designed and manufactured by laser-writing. It is formed by a two-dimensional arrangement of 23 channel waveguides in a "zig-zag" configuration. 
Such a configuration of channel waveguides fulfils simultaneously the condition for interferometric beam combination that a simple linear array would not support, and allows the spectral dispersion of the signal on the focal plane array, as shown in Diener et al. (2016)\cite{Diener2016}\,. 
\begin{figure}[t]
\centering
\includegraphics[width=1\textwidth]{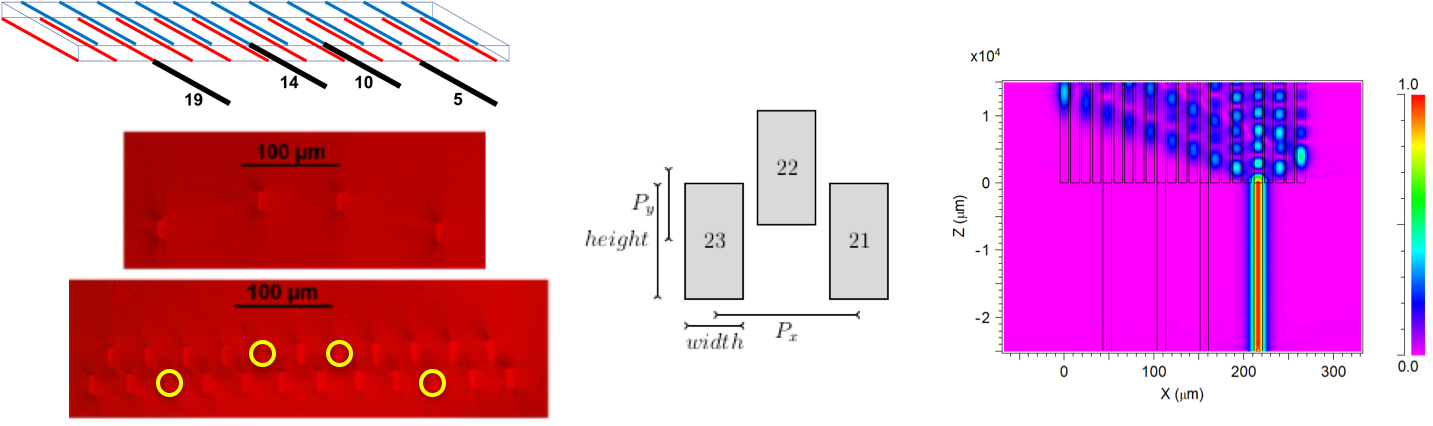}
\caption{Left: principle of the zig-zag arrangement of the DBC with its feeding waveguides. A microscope view of the input and output is shown. Middle: example of a possible arrangement of the DBC geometry. Right: Rsoft simulation showing the coupling to the nearby channel waveguides over the propagation distance when flux is injected in one input (simulations by Th. Poletti).}\label{dbcfab}
\vspace{0.0cm}
\end{figure}
Each of the 23 waveguides were written using the multipass technique, leading to cross-section of $\sim$25$\times$10\,$\mu$m and separated from each other by $\sim$16--30\,$\mu$m (see Fig.~\ref{dbcfab}). In this case, the 3D capabilities of the ULI platform are of great importance to fabricate this two-dimensional lattice. This would not be possible with planar lithography. 
Prior to the 23-waveguide portion of the DBC, a "pre-feeding" section is written in order to directly couple the telescope beams into the appropriate subset of four waveguides. 
Two DBC components were written: DBC\#\,1 had a total length of 5\,cm (2.5\,cm for the pre-feeding section and 2.5\,cm for the DBC section);  DBC\#\,2 had a total length of $\sim$2.5\,cm (1\,cm for the pre-feeding section and 1.5\,cm for the DBC section). For both samples, a mitigation of the long range tensile stress was applied\cite{Diener2018}. 
The light injected into one input waveguide is coupled out to all the 23 waveguides after a sufficient propagation distance. When all four inputs are injected simultaneously, the resulting output interference pattern imaged onto the camera can be related to the source coherence function via the P2VM matrix (see Sect.~\ref{v2pm}). 
The robustness of the approach has been demonstrated at visible wavelengths for spectro-interferometry\cite{Saviauk2013}\, for three telescopes, and first results have in the mid-infrared have been published in Diener et al.(2017)\cite{Diener2017}\,. 




\section{Characterization and laboratory tests}\label{Labwork}

Laboratory testing and characterization in ALSI has been carried on towards two directions:  on one side we have concentrated on the understanding of the physical processes involved in the ULI technique and their potential impact on the optical properties of the written structures (cf. Sect.\ref{DeviceFab}), and on the other side in establishing the monochromatic and polychromatic interferometric performances of new mid-infrared devices based on metrics relevant for an astronomical usage. 
In this section we summarise some important results that are presented in more details in two companion papers in these proceedings, Diener et al. "Beam combination schemes and technologies for the Planet Formation Imager" and Tepper et al. "Photonics-based mid-infrared interferometry: the challenges of polychromatic operation and comparative performances". 

\subsection{2-telescope evanescent couplers}

\begin{figure}[t]
\centering
\includegraphics[width=0.8\textwidth]{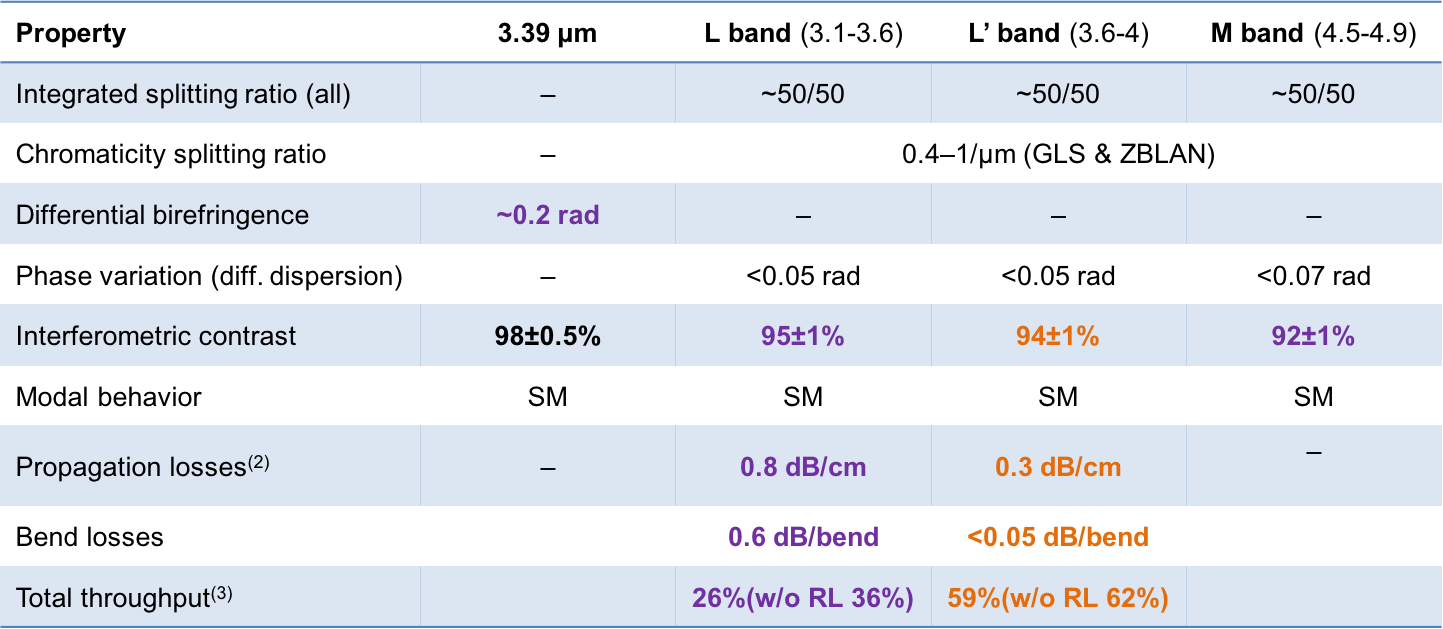}
\caption{Performance overview for directional couplers based on GLS and ZBLAN glasses. Orange = ZBLAN glass; Pink = GLS glass; (1) Variance of the contrast limited by non simultaneous photometric correction; (2) Intrinsic + extrinsic (impurities) losses;
(3) Fresnel losses (or RL for reflection losses): GLS glass (~30\%), ZBLAN (~7\%).}\label{summary}
\vspace{0.0cm}
\end{figure}

\begin{figure}[t]
\centering
\includegraphics[width=0.9\textwidth]{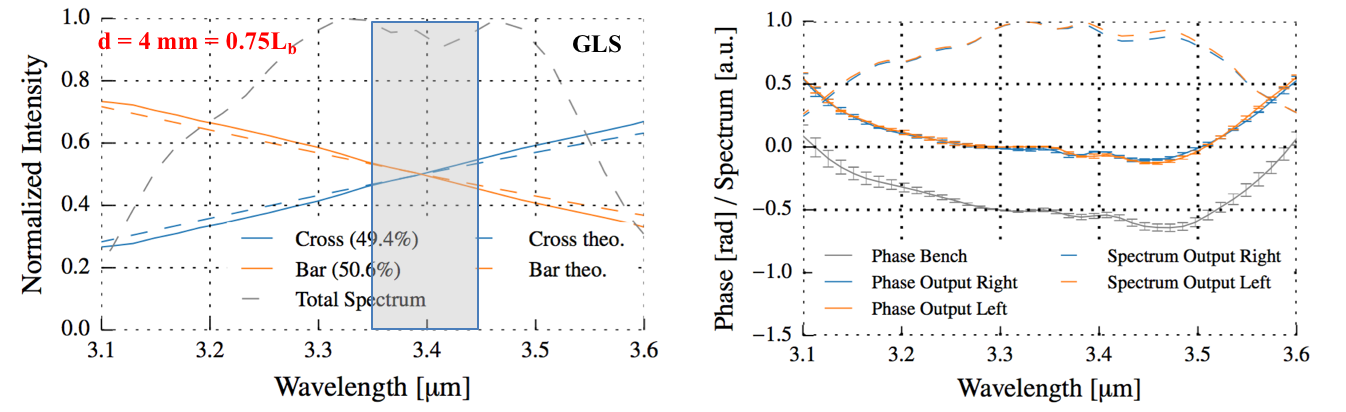}
\caption{Left: splitting ratio of the GLS directional coupler as a function of $\lambda$\cite{Tepper2017a}. The slope is measured to $\sim$1\,$\mu$m$^{-1}$. For the ZBLAN coupler (not shown here) the slope is shallower and $\sim$0.4\,$\mu$m$^{-1}$. The grey area shows the bandwidth over which the equivalent extinction has been calculated. Right: Phases (blue and red) for the two chip outputs over the L band. The value $\pi$ was subtracted from one of the chip outputs, and it is clearly visible that the expected $\pi$ phase shift is almost perfectly achromatic.
}\label{splitratio}
\vspace{0.0cm}
\end{figure}

The simple evanescent coupler has been studied in details in ALSI with components written and tested both in ZBLAN and GLS glasses. This exhaustive characterisation has permitted to assess that the ULI platform is performing well in both the aforementioned substrates. The overall throughput is currently found to be higher in the fluoride-based component than in the Chalcogenide-based one. Fig.~\ref{summary} summarises the most important results. Details on the geometry of the couplers can be found in Tepper et al. (2017a, 2017b)\cite{Tepper2017a,Tepper2017b}. 
An interesting aspect on the directional coupler regards the level chromaticity of the splitting ratio across the band. As expected for the directional coupler, the flux splitting at the each output is found to depend on the wavelength with a more or less marked slope (see Fig.~\ref{splitratio}). While a chromatic splitting is acceptable for $V^2$ interferometry, this is potentially more penalising for nulling interferometry which typically aims at operating over a broad spectral range to maximise the sensitivity. \\ 
For the GLS and ZBLAN couplers, the spectral slopes for unpolarized light are found to be 1\,$\mu$m$^{-1}$ and  0.4\,$\mu$m$^{-1}$, respectively. By reducing the bandwidth to 100\,nm (3\% bandwidth) and 200\,nm (6\% bandwidth) around 3.4\,$\mu$m and 3.3\,$\mu$m for the GLS and ZBLAN, respectively, and assuming the ideal case where the coupler is not affected by differential birefringence or dispersion, the extinction at the nulled output of the directional coupler would still go down to $<$10$^{-3}$, which would be close to the requirements of the Hi5 instrument\cite{Defrere2018}. 
Here the extinction is given by $\rho$=(1-$V$)/(1+$V$). 
However, this conclusion should be taken with caution since only a dedicated lab nulling experiment can corroborate this result. Nonetheless, this points out that cascaded evanescent couplers might still be successfully implemented for ground-based nulling\cite{Errmann2015}.\\
\\
For both couplers, the two interferometric outputs are found to be in phase opposition at all wavelengths with great accuracy, which illustrates the good behaviour of the device. This can be well seen in Fig.~\ref{splitratio}. \\
\\
In terms of polarization behavior, the splitting ratio of the GLS directional couplers seems to show some dependence on the input polarisation state, but within 10\% of the 50/50 splitting ratio when the input polarisation is varied from 0$^{\circ}$ to 180$^{\circ}$.\\
On the other side, the characterisation of the  asymmetric couplers presented in Sect.~\ref{dircouplers} did not resulted yet in a significantly flatter splitting ratio across the L, L$^{\prime}$ bands. \\
\\
Last but not least, we assessed that no impact on the coupler performances was observable following a 10 hours cooling-cycle down to 120K for the component.

\subsection{2-telescope ABCD beam combiner}

The ABCD beam combiner developed within ALSI was deigned to operate around at 3.4\,$\mu$m. The monochromatic characterisation with the 3.39\,$\mu$m linearly polarised laser shows that the four outputs are reasonably found to be in quadrature (see Fig.~\ref{abcd}). However, as we deviate from this design wavelength, first results on the spectral analysis show that the fringe quadrature is not maintained. 
\begin{figure}[b]
\centering
\includegraphics[width=0.85\textwidth]{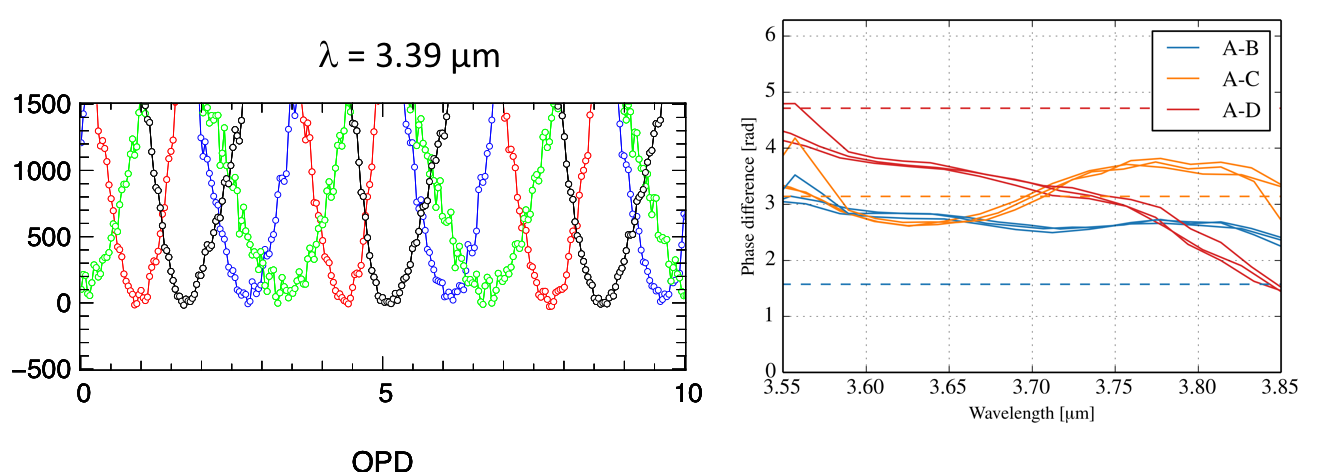}
\caption{Left: monochromatic and linear-polarization interferograms measured at the ABCD outputs. The different sine waves should ideally be separated by $\lambda$/2 from each other. Right: phase differences in the ABCD as a function of the wavelength. The dashed lines indicate the expected location of the phase curves for the different ABCD outputs. It appears that the inserted $\pi$/2 phase shift is significantly chromatic far from 3.39\,$\mu$m. }\label{abcd}
\vspace{0.0cm}
\end{figure}
Furthermore, probably because of a higher number of bend waveguides involved in this design, the total throughput of the component is less than 10\% as it is probably dominated by the bending losses. A detailed characterisation of the monochromatic V2PM of the 2T-ABCD device is reported in Diener (2017)\cite{Diener2017}.\\
Further characterisation on new laser-written components is needed in order to further assess how compliant the ULI platform can be with the instrumental requirements of an ABCD beam combiner. 

\subsection{4-telescope Discrete Beam Combiner}

The DBC concept presented in Section 3 was tested in monochromatic ($\lambda$=3.39\,$\mu$m) and polychromatic (L$^{\prime}$ band) light using two different components with different length, but both undergoing tensile stress compensation. The DBC\,\#1, that measured circa 5\,cm long, was only tested monochromatically. The DBC\,\#2 had a total lenght of circa 2.5\,cm with a combination section of circa 1.5\,cm. For this component preliminary tests were conducted in broadband light and dispersed mode thanks to a grating\footnote{Thorlabs Item \#GR1325-30035}. We also performed R-soft simulations at a design wavelength of 3.4\,$\mu$m to better understand the properties of the DBC's V2PM matrix and its theoretical propagation losses.

\subsubsection{Simulation analysis}

\begin{figure}[t]
\centering
\includegraphics[width=0.54\textwidth]{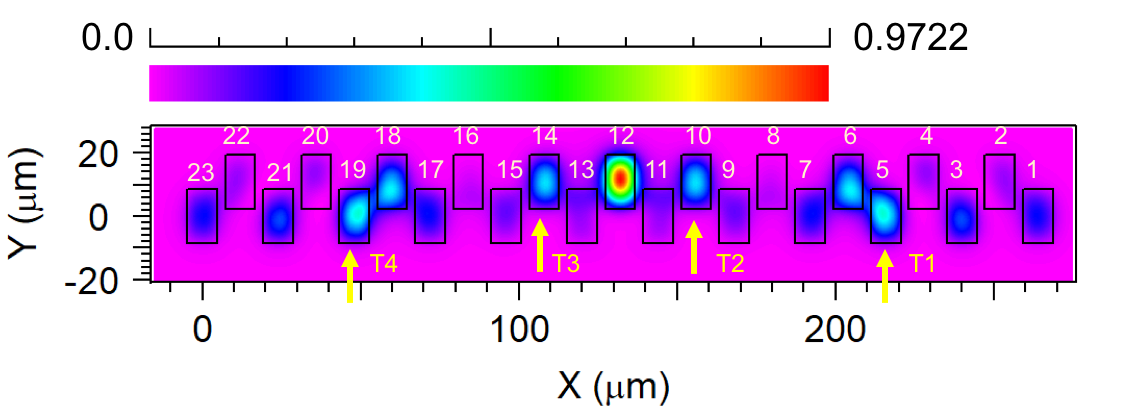}
\includegraphics[width=0.45\textwidth]{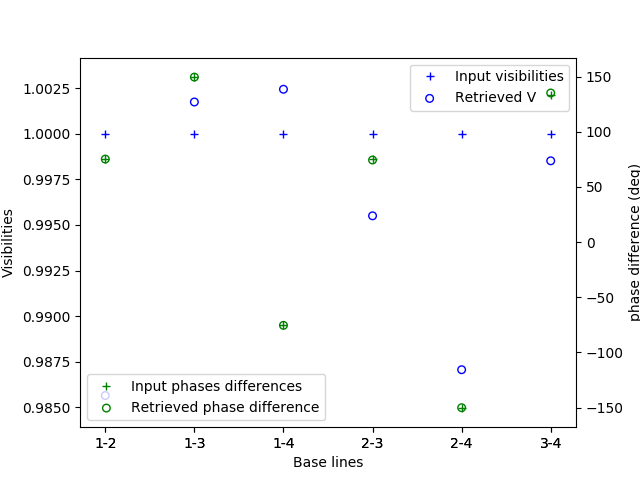}
\caption{Left: RSoft simulation of the DBC output when all the four telescope beams are coupled. 
Right: test of the retrieval of the coherence function with the DBC using a 4-telescope input vector with $P_1$=1, $P_2$=2, $P_3$=3, $P_4$=4, $V$=1, and $\phi_{1,2,3,4}$=[75$^{\circ}$, 150$^{\circ}$, 225$^{\circ}$, 0$^{\circ}$].}\label{dbcrsoft}
\vspace{0.0cm}
\end{figure}

The layout of the DBC output is shown in Fig.~\ref{dbcrsoft}. The 23 rectangular outputs are marked in black. The four inputs are marked with an arrow and correspond to a canonical case with a power of unity per input and zero phase-shift between them. The index contrast is $\Delta$n=5$\times$10$^{-3}$.
For each of the 6 baselines, the terms $V^{inst,j}_{lm}$ and $\phi^{inst,j}_{lm}$ of Eq.~\ref{v2pmeq} were retrieved by injecting two telescopes at a time, scanning a phase ramp and fitting the recorded interferogram at each of the 23 outputs. The monochromatic instrumental visibilities are high for all baselines, though the contrast value depends on how the overlap integral is calculated at each of the 23 outputs. 
Similarly, the 23$\times$4 $\kappa$-matrix terms $t_{kj}$ are derived by injecting one photometry at a time. This permits to derive the simulated V2PM linking the source coherence function to the measurable intensity vector {\bf M}. 
We derive a condition number of CN$\sim$7.3. A coarse physical interpretation of this number is the ratio of the errors between the input and the output vectors\footnote{The V2PM and P2VM exhibit the same condition number CN}. 
Testing the retrieval of an arbitrary coherence function by mean of the pseudo-inverse P2VM showed an average relative deviation of 0.5\% for the visibilities and 0.3\% for the phase differences. This primarily results from the limited accuracy with which the 23 output intensities $M_i$ are retrieved due to some level of cross-talk between the DBC outputs. 
By increasing the index contrast from 5$\times$10$^{-3}$ to 7$\times$10$^{-3}$, the confinement of the field is improved but only a negligible decrease in the CN is observed from 7.3 to 7.1.
\\
The theoretical throughput of the component is estimated by exciting one input by a unity power and computing the sum of the 23 output fluxes\footnote{The output flux of one waveguide is estimated through the overlap integral between the waveguide mode and the mode of an elliptical fiber}\,. The relatively good field confinement in the waveguide leads to a throughput of $\sim$87\%, with little energy actually lost in the cladding. This value also includes some level of coupling losses that are not estimated here. Assuming they are negligible, this would result in 0.12\,dB.cm$^{-1}$ propagation losses or better for a 5-cm long component. 

\subsubsection{Experimental characterisation in monochromatic light}

The DBC\,\#1 was characterized at 3.39\,$\mu$m by measuring successively the 23 instrumental visibilities $V_{lm}^{inst,j}$ and phases $\phi_{lm}^{inst,j}$ for each baseline $lm$, as reported in Diener\,et\,al.\,(2017)\cite{Diener2017}\,. The high monochromatic instrumental contrasts close to 90\% point to a low-level of differential birefringence, which would be typically the first cause for a decrease of the photometrically calibrated instrumental contrast. Note that the test was performed with linearly polarised light. 
The experimental retrieval of the coherence function for a monochromatic point-like source shows that a visibility of one with an absolute error of 5\% to 10\% is obtained. The phase information can also be properly retrieved with a standard deviation of $\sim$0.16\,radis ($\sim$9$^\circ$) on the residual. The monochromatic characterisation of the DBC\,,\#2 was, at the contrary, unsuccessful as it was not possible to observe flux at all the 23 outputs, whatever the launched input beam. This may point, for unknown reasons, either to a problem linked to the manufacturing process and its repeatability, or to a design issue. Indeed, when the same component was tested around 3.8\,$\mu$m, all the 23 outputs were properly observed (see Fig.~\ref{slopedbc}).

\subsubsection{Experimental characterisation in polychromatic light}

\begin{figure}[t]
\centering
\includegraphics[width=0.24\textwidth]{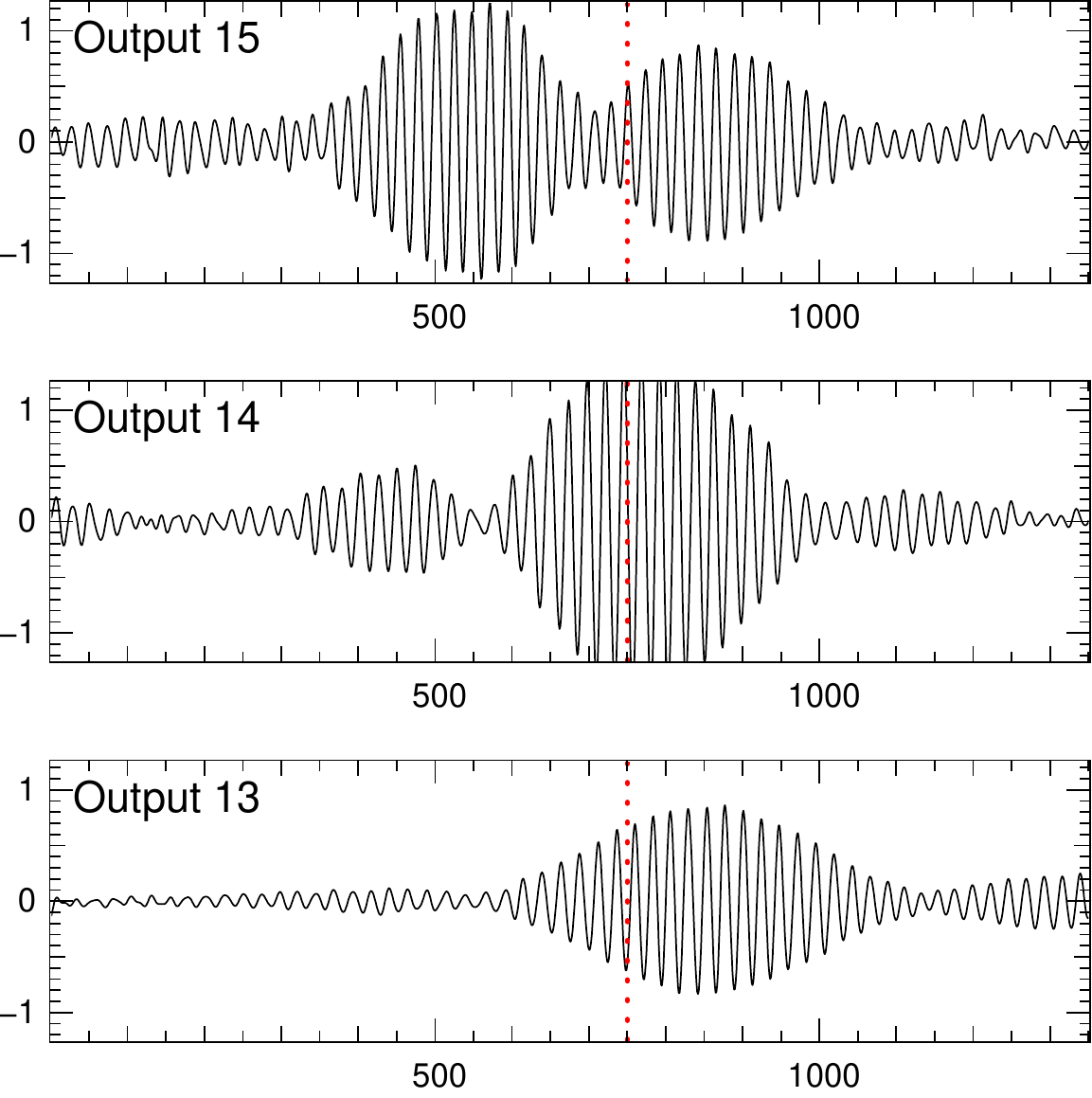}
\includegraphics[width=0.24\textwidth]{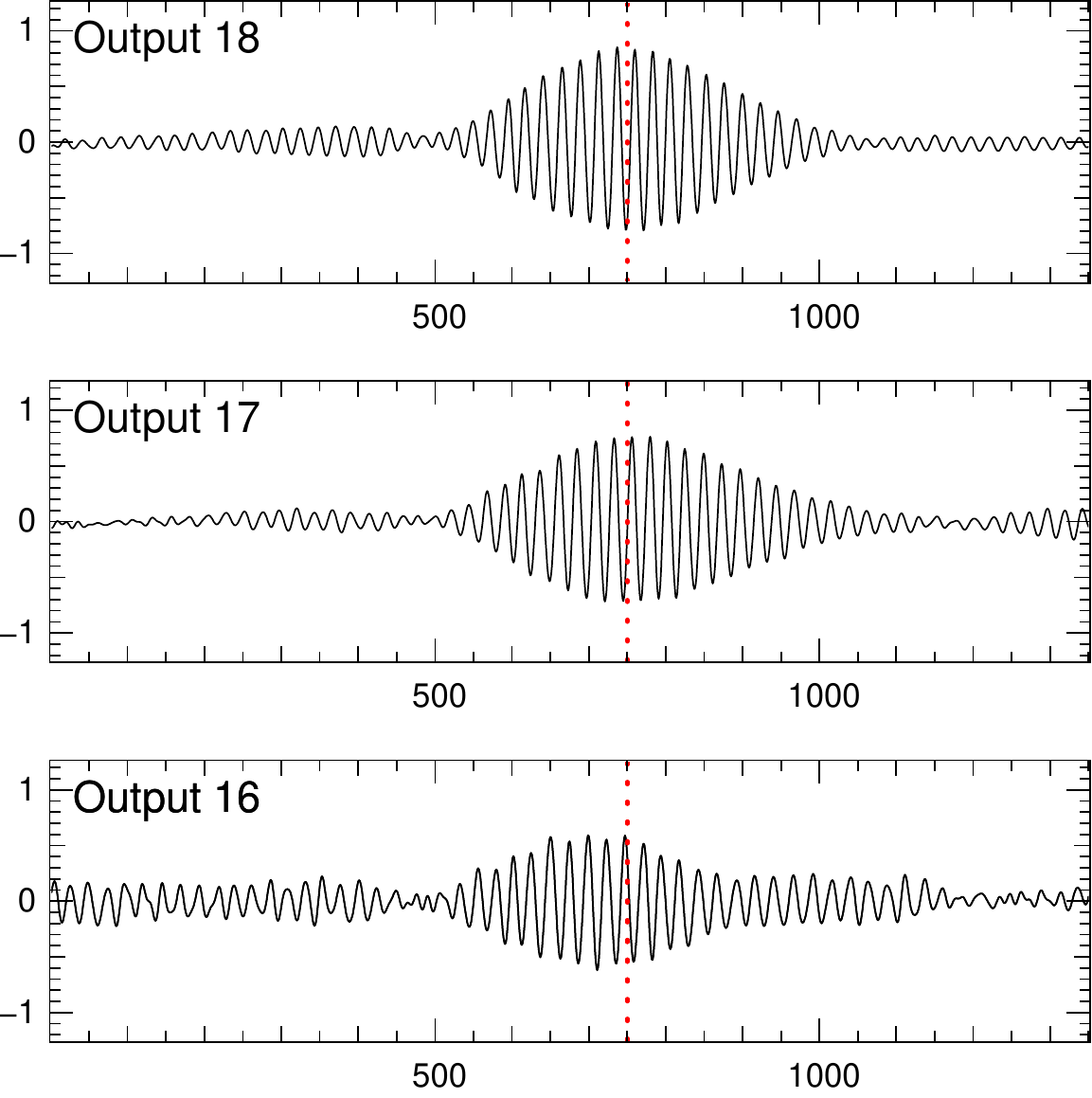}
\includegraphics[width=0.24\textwidth]{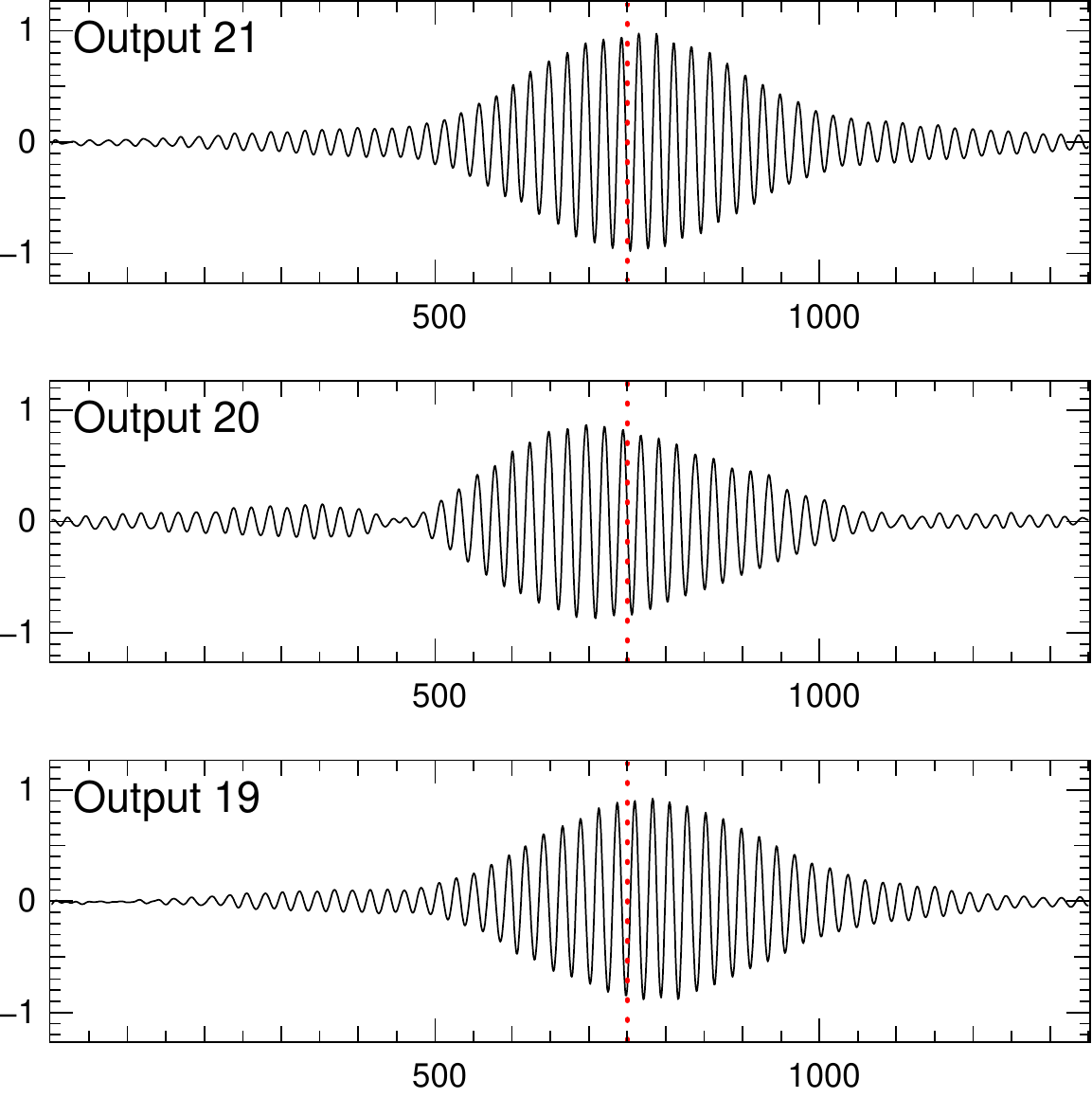}
\includegraphics[width=0.24\textwidth]{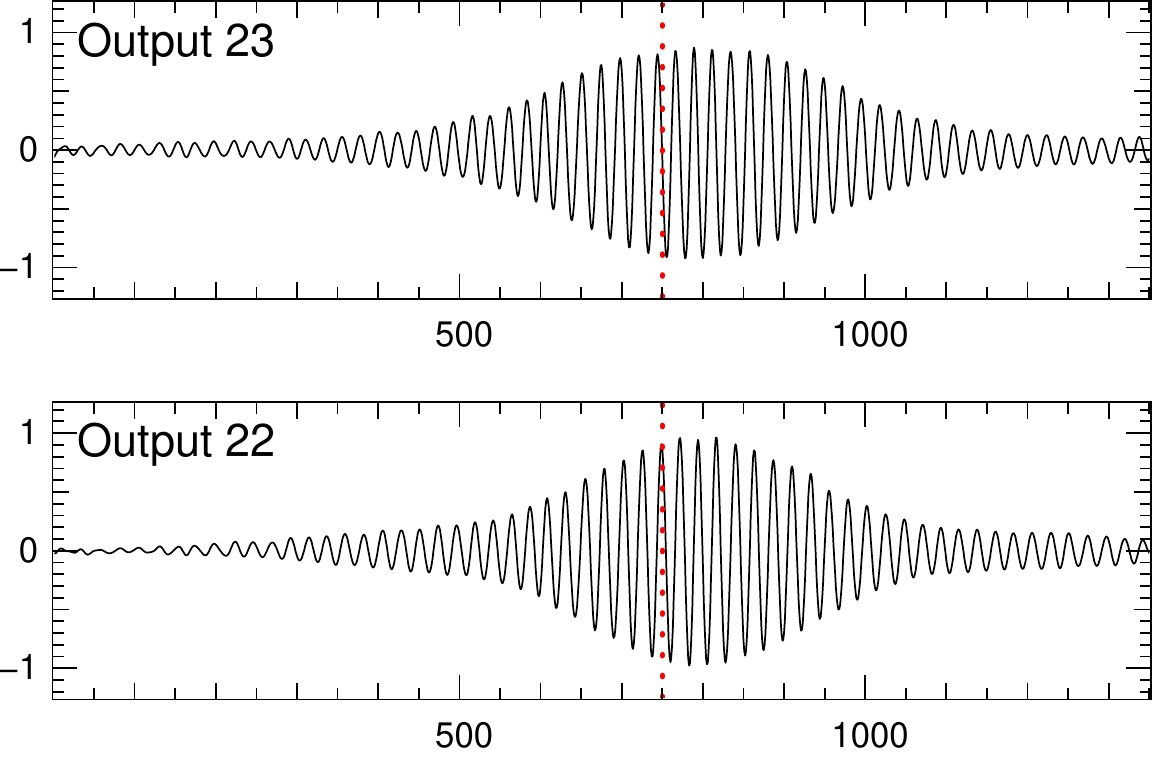}\\
\includegraphics[width=0.24\textwidth]{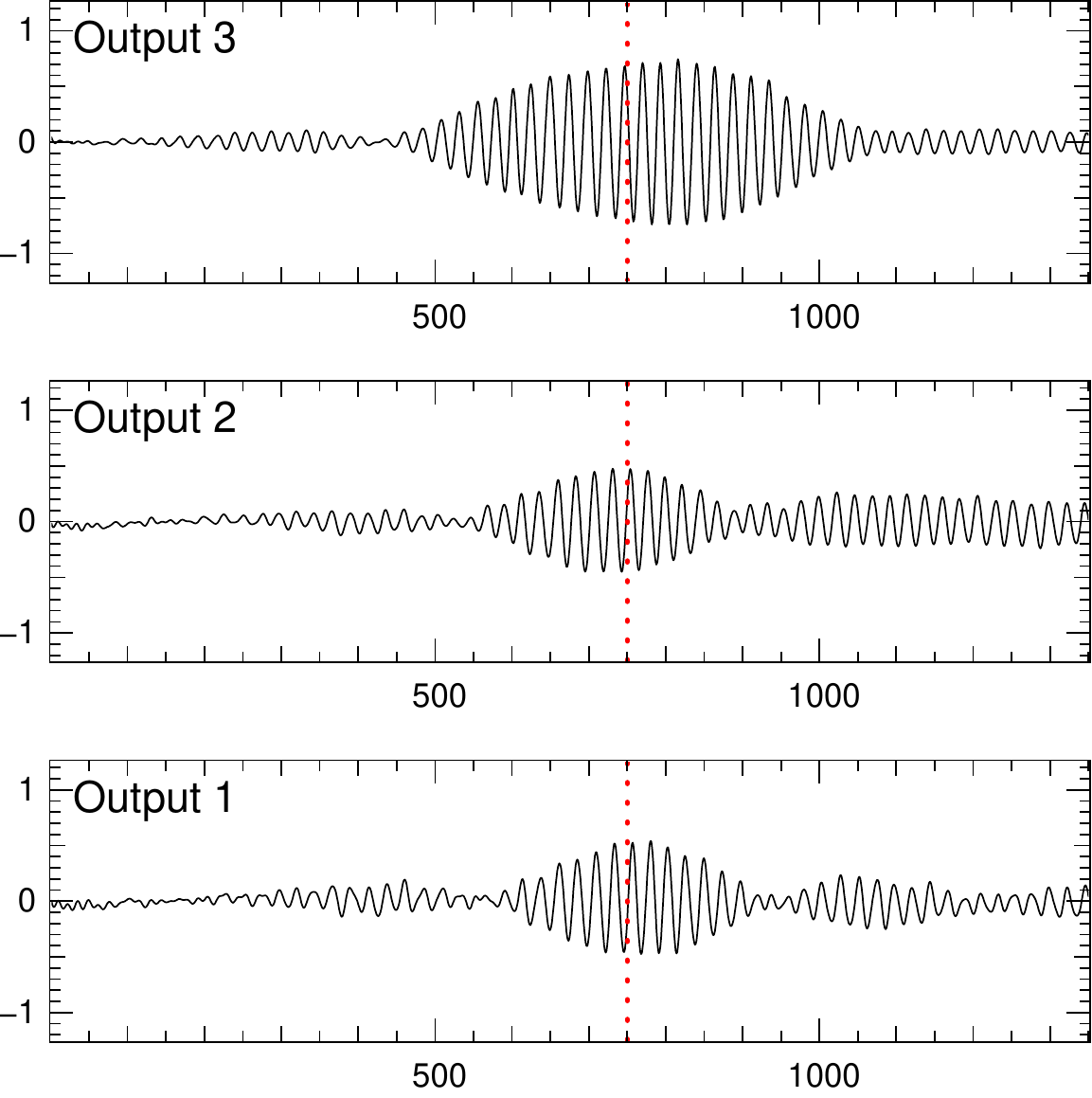}
\includegraphics[width=0.24\textwidth]{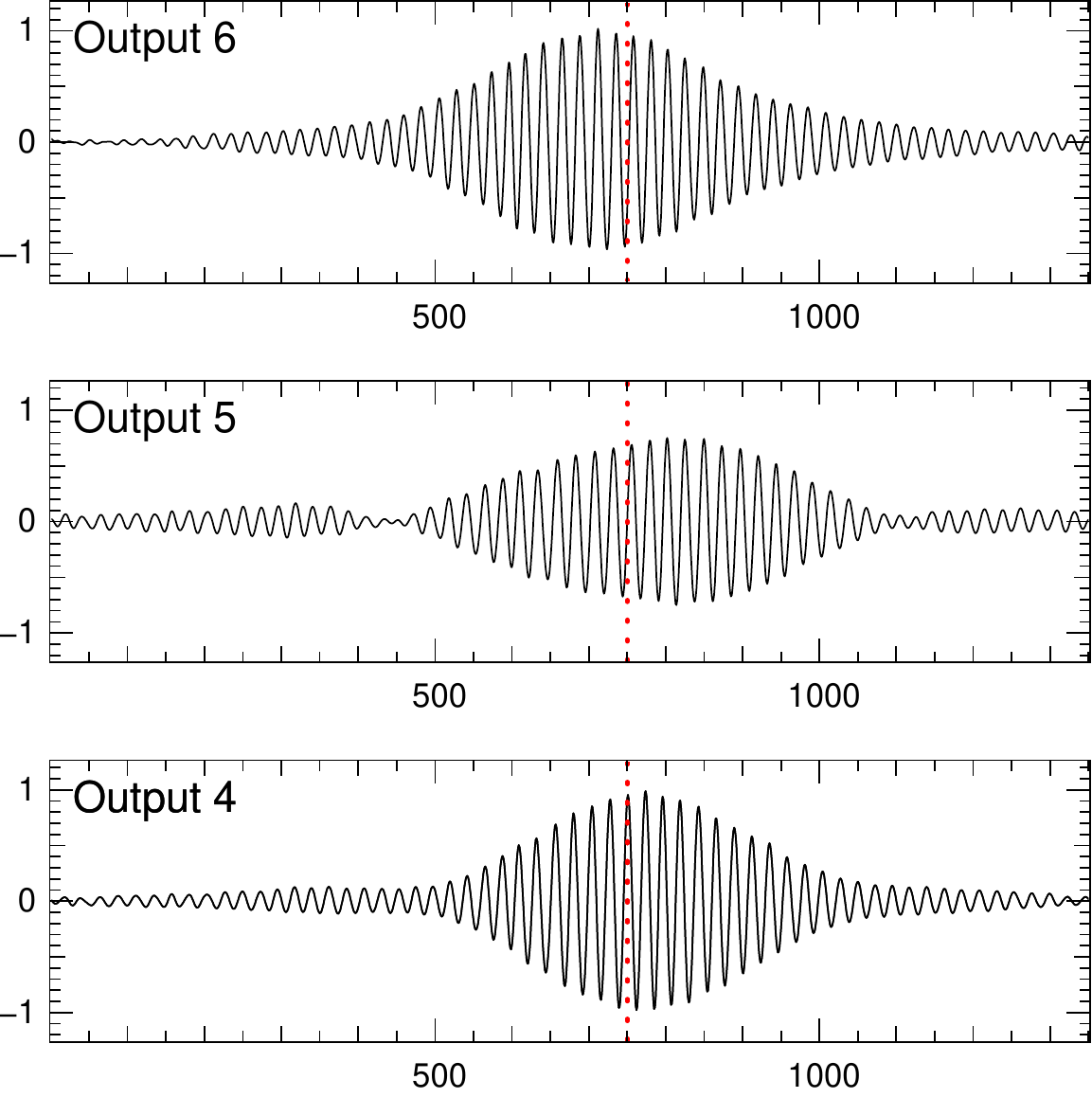}
\includegraphics[width=0.24\textwidth]{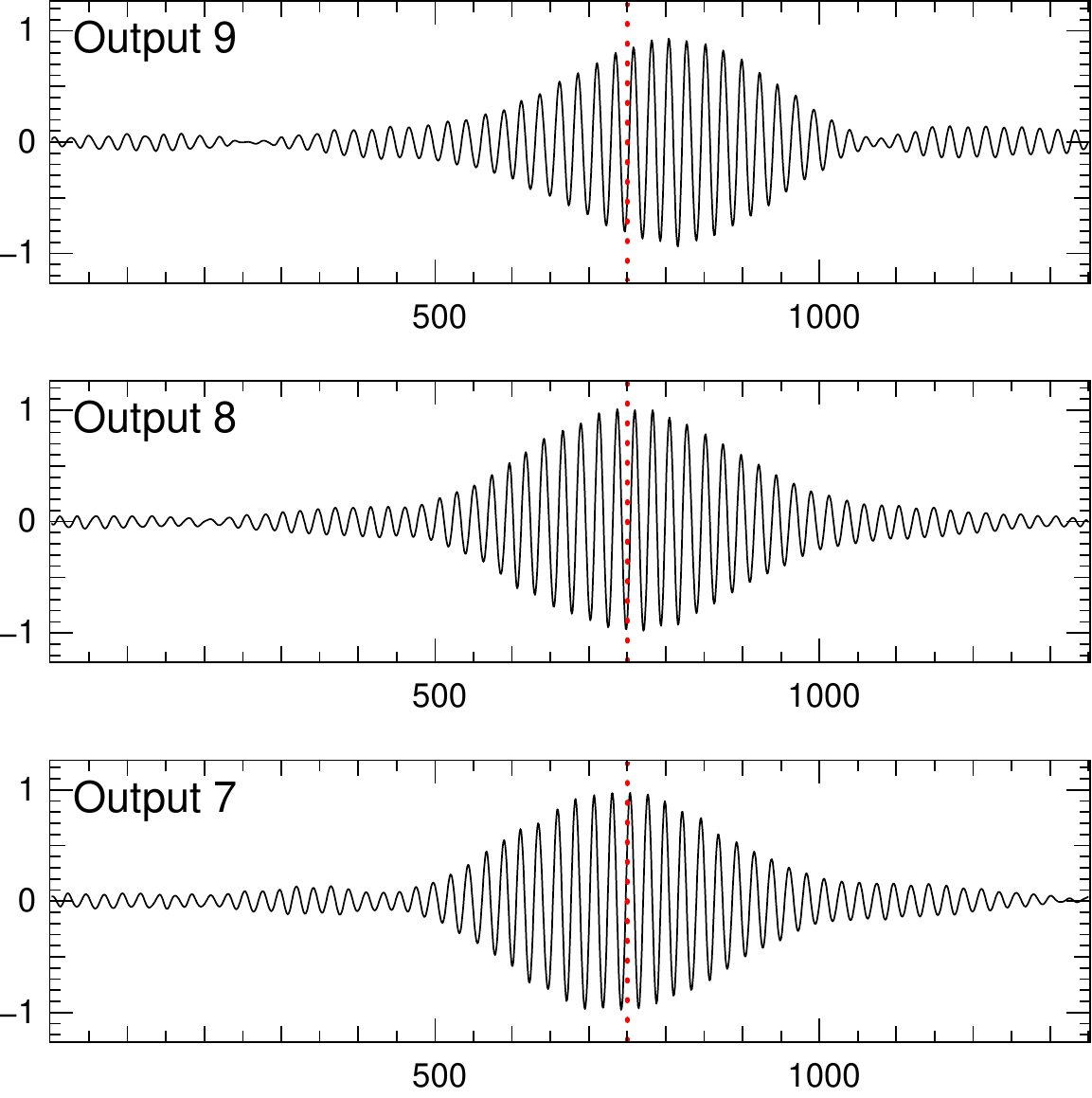}
\includegraphics[width=0.24\textwidth]{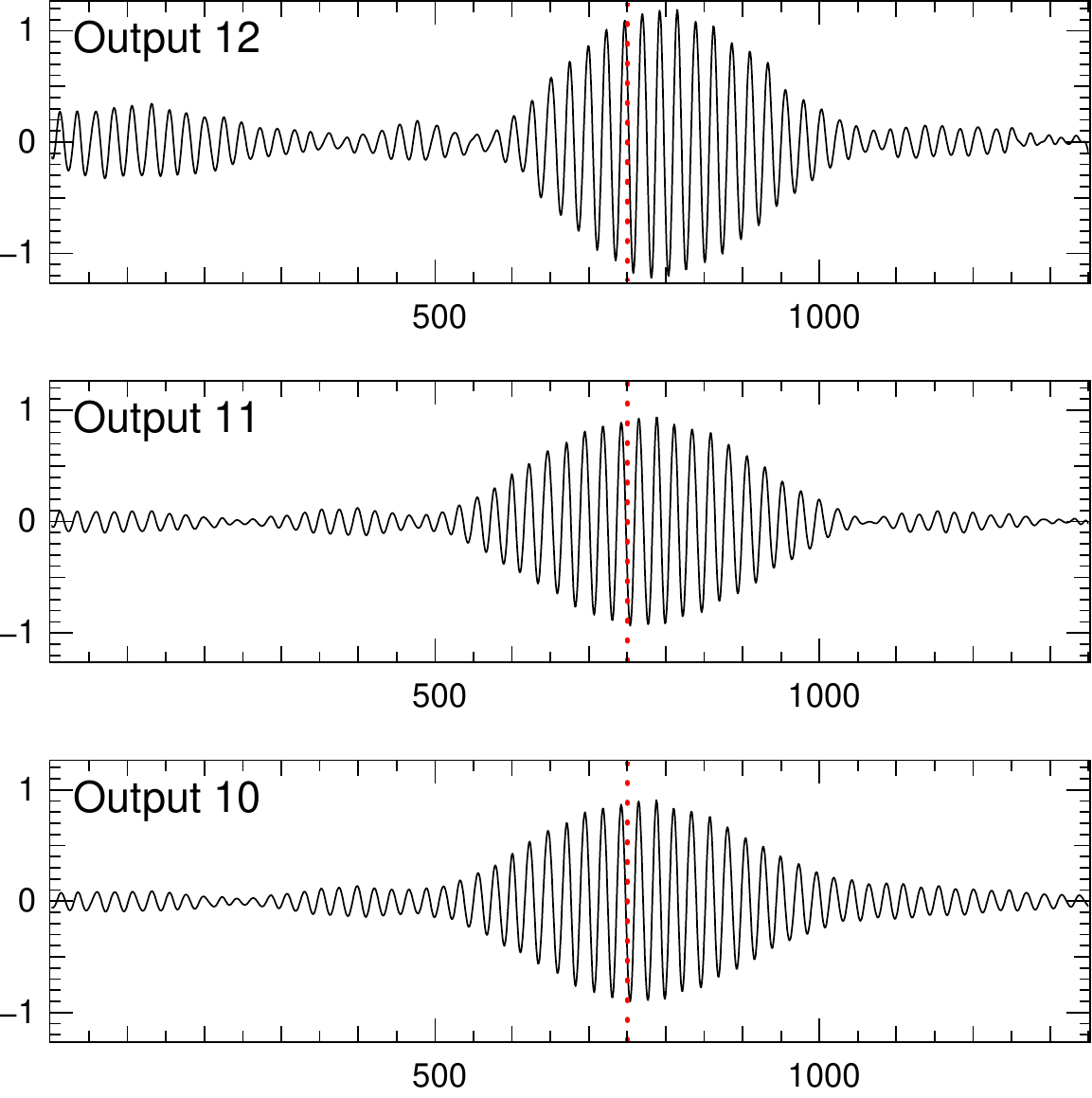}
\caption{Whitelight interferograms observed at the 23 DBC outputs for the baseline 2-3 (waveguides \#10 and \#14)}\label{dbcout_b23}
\vspace{0.0cm}
\end{figure}

\begin{figure}[t]
\centering
\includegraphics[width=0.54\textwidth]{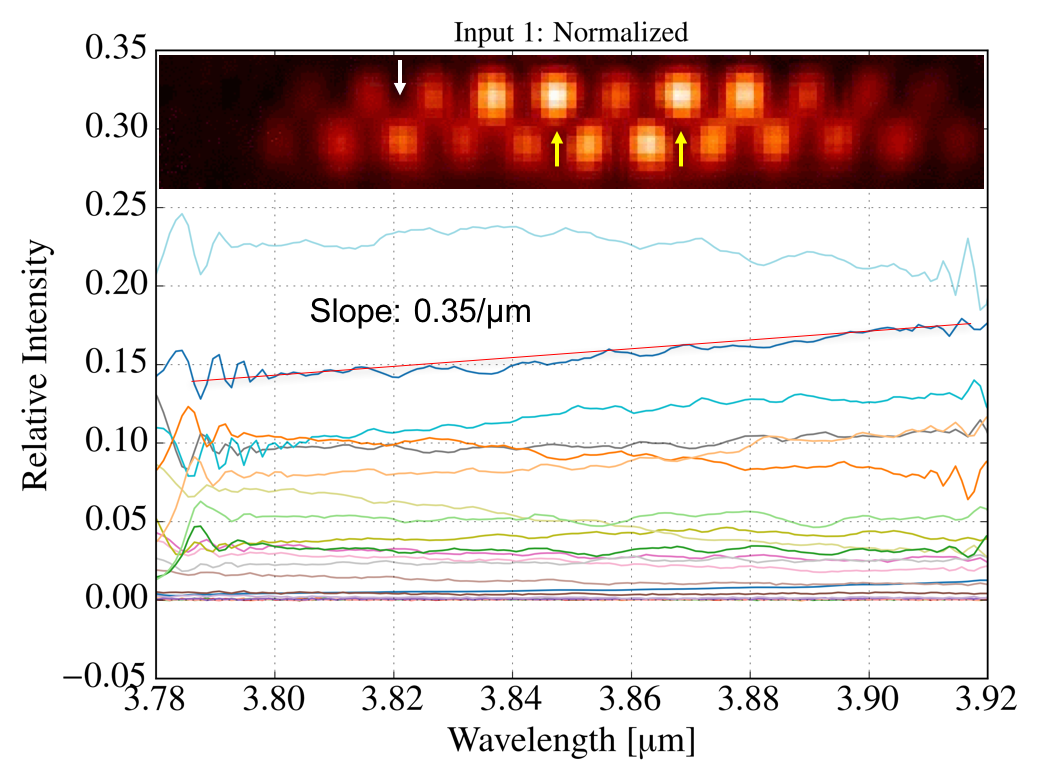}
\caption{FTS spectra of the 23 DBC outputs normalised to the total spectrum when the input indicated by the white row is injected.  These curves give an indication of the chromaticity of the splitting ratio, which is found to be less than 0.35\,$\mu$m$^{-1}$. The top view shows all the 23 DBC outputs when the baseline 2-3 (waveguides \#10 and \#14) is measured.}\label{slopedbc}
\vspace{0.0cm}
\end{figure}

Saviauk et al. (2013)\cite{Saviauk2013} tested the operation of the DBC in visible polychromatic light coupled to a dispersing element. It was shown the condition number of the V2PM increases with an increasing bandwidth of a spectral channel. At the contrary, for the 4-telescope ABCD-type beam combiner, it was shown that it can operate in broadband conditions\cite{Jocou2014}\,, with the undispersed white-light interferograms exhibiting the expected nominal 0, $\pi$/2, $\pi$, 3$\pi$/2 phase shifts at the A, B, C and D outputs. This is due to the intrinsic symmetry of the ABCD beam combiner with respect to a given input baseline.\\
In the case of the DBC combiner, since the coupling strength between the different waveguides depends on the wavelength, it is expected to deliver poor performances in terms of contrast when operating in broadband conditions. For those output waveguides positioned in rather asymmetric arrangement with respect to the input baseline (e.g. the outputs \#1 or \#23 with respect to the input baseline composed of the waveguides \#5 and \#19 in Fig. 4), the term of differential chromatic dispersion might be dominated by $\beta\Delta L$. \\
At the time of writing, the interferometric and V2PM characterisation in dispersed light was not completed. Meanwhile, we analysed the retrieved interferometric outputs obtained in polychromatic and linearly polarised light with a bandwidth of 300\,nm around 3.8\,$\mu$m. Clearly, this test cannot be definitive since we operate far from the design wavelength and in undispersed mode. In Fig.~\ref{dbcout_b23} are shown the measured photometrically corrected white-light interferograms for the central baseline composed by the waveguides \#10 and \#14. We measure interferometric contrasts between 40\% and 90\%, and the interferograms do not show on average significant dispersion. For other baselines, e.g. 1-4 (waveguides \#5 and \#19), lower contrasts of $\sim$25\% are measured.\\
Similarly to the case of the directional couplers, we also looked at the chromaticity of the DBC splitting ratio by performing FTS spectroscopy of each of the 23 outputs normalised to the total transmitted spectrum (i.e. sum of all the 23 spectra). Interestingly, and despite this measurement was limited to a bandwidth of 100\,nm, we observed that the slope of the splitting ratio was less than $\sim$0.35\,$\mu$m$^{-1}$, i.e. close to the best result obtained with the ZBLAN coupler (see Fig.~\ref{slopedbc}). We believe this is an important advantage to be considered if the DBC is to be used to deliver high-precision visibilities with moderate spectral dispersion. \\
Finally, a full characterisation of the polychromatic was not conducted as for a number of baselines a proper measurement of the instrumental visibility and phase was not retrieved for a number of outputs due to low signal-to-noise ratio. Further results will be obtained in the near future when operating in dispersed mode.

\subsubsection{Throughput of the DBC in polychromatic light}

A further -- and important -- test on the DBC component was its total throughput in the L$^{\prime}$ filter. The advantage of the DBC component over other architectures is that only channel waveguide are employed, hence avoiding bending losses. \\
We measured the total throughput -- including Fresnel losses -- of the DBC by injecting the four input beams one at a time and found a transmission as high as 60\%, 50\%, 51\% and 43\%. Considering the architecture of the DBC, we believe this is a very promising result when considering the total throughput as a top requirement for on-sky interferometry.\\
Though the implementation of a DBC supposes to "split" the light from one telescope into 23 waveguides, its efficiency is not necessarily inferior to the 4T-ABCD combiner, which is composed of 4 input and 24 output waveguides. Going to six telescopes, the DBC concept can be operated with 41 output waveguides (see Pedretti et al., these proceedings), whereas the 6T-ABCD will employ -- at least -- 60 output waveguides to perform the static ABCD encoding.










\section{Perspectives and conclusions}

The ALSI project has successfully paved the way for an astronomical use of mid-infrared integrated solutions based on the 3D ultrafast laser writing technique. We have shown in details that performances compatible with a future use at the telescope could be reached. A number of technical issues still remain to be properly addressed, such how to obtained higher confinement and lower bending losses in ABCD components, but the prospects are promising. The simple DBC architecture can be implemented in a relatively straightforward manner in mid-infrared substrates such as GLS.\\
In the future, new achromatic directional couplers will be investigated (for instance with interaction lengths equal to half a beat-length instead of a quarter beat-length), considering cascaded architectures. For the DBC, the addition of a fan-out section in order to better isolate the 23 outputs should also ease the instrumental implementation and consequent data processing. The on-going NAIR project\footnote{https://www.lsw.uni-heidelberg.de/projects/NAIR/cmsms/} will investigate some of these aspects at shorter wavelength, e.g. for the K\,band. 




\acknowledgments          
LL and SM acknowledge financial support by the BMBF through the grants 05A14PK2 and 05A14SJA "ALSI -- Advanced Laser writing for Stellar Interferometry", as well as by the DFG through the grant 326946494 "NAIR -- Innovative Astronomische Instrumentierung mittels photonischer Reformatierer". 

\bibliography{report}   
\bibliographystyle{spiebib}   

\end{document}